\documentclass{ptapap}

\author{John D Landstreet}[UWO,AOP]

\affil[UWO]{Department of Physics \& Astronomy, University of Western Ontario, London, ON, Canada N6G 1P7}
\affil[AOP]{Armagh Observatory \& Planetarium, College hill, Armagh, BT61 9DG, Northern Ireland}

\title{Reflections on the discovery of the first magnetic white dwarf}

\begin{document}

\maketitle

\begin{abstract}
I was one of the six people most closely involved in the discovery of
the first magnetic white dwarf in 1970, now 50 years ago. Thinking
back on this event, I have realised that the discovery occurred when
and how it did because of a series of lucky coincidences along a
strange and winding path. In this
paper I recount the events as I recall them, and reflect on how those 
unlikely coincidences helped us to succeed.

\end{abstract}

\section{Background}

In 1947, Horace Babcock announced the detection of a magnetic field on
the surface of a main sequence A-type star \citep{Babc47}. This was
the first detection of a field in any star other than that of the Sun,
whose magnetic field had already been studied for some 30 years. This
discovery aroused a lot of interest among astronomers. Almost
immediately, P. M. S. Blackett proposed a theory suggesting that the
magnetic moment of a planet or a star should be proportional to its
angular momentum \citep{Blac47}. In particular, he suggested that
strong magnetic fields might exist in white dwarf stars, which
Blackett assumed would be very rapid rotators, although at the time
nothing was actually known observationally about rotation rates of
white dwarfs. Blackett discussed detection of stellar surface fields
by the Zeeman effect, and he suggested that unidentified weak
absorption features found in the star Grw+70$^\circ$\,8247 might be
Zeeman patterns of H$\gamma$ and H$\delta$, broadened by a field of
the order of $10^7$\,Gauss.

By the early 1960s, Babcock and others had established that magnetic
fields occur on the surfaces of members of a specific class of upper
main sequence A and late B stars, the ``peculiar A'' (Ap) stars
\citep{Babc58a,Babc58b}. These stars are in fact quite {\em slow}
rotators, and so Blackett's ideas fell out of favour. 

The fields found in the Ap stars were shown to have strengths
generally of the order of hundreds or thousands of Gauss, to vary
periodically as the host star rotates and shows the field from
different directions, but to display negligible secular change. The
fields were found to be roughly dipolar in global structure. The
various observed magnetic and related (spectrum and phometric
variability) phenomena were satisfactorily unified and explained in
the ``oblique rotator'' model \citep{Stib50}.

However, the origin and nature of these static fields remained a
puzzle. In the context of an effort to explain X-rays produced by the
Crab Nebula, a radically different theory of occurrence of magnetic
fields in stars was proposed by Lodewijk Woltjer, namely that magnetic
flux might be roughly conserved during stellar evolution
\citep{Wolt64}.  Woltjer pointed out that the Crab might host one of
the hypothetical neutron stars, and that such an object might posess a
huge magnetic field.

Woltjer estimated the possible field strength from flux
conservation, as 
\begin{equation}
B(R) \approx B_0 (R_0/R)^2 \approx (\rho/\rho_0)^{2/3}
\end{equation}
where $R$ is the stellar radius, $\rho$ is the mean stellar density,
and 0 subscripts refer to the initial (or at least earlier)
state. Assuming that the initial star had been a main sequence staar
with an internal field of $B_0 \sim 10^4$\,G or so and a radius of
about $10^{6}$\,km, which finally collapsed to a radius of order
10\,km, the resulting neutron star might have a field of the order of
$10^{14}$\,G

It was at this point that I entered the picture. I had arrived back in
New York City at Columbia University in 1962 as a graduate student in
the Physics Department. In March 1964, I started work on a thesis
under the supervison of Woltjer, who had recently been appointed
appointed chair of the Columbia Astronomy Department. The basic
premise of the thesis topic was an obvious extension of Woltjer's idea
about flux conservation. He pointed out that the same reasoning as for
neutron stars suggested that there might be magnetic fields of the
order of $10^8$\,G in some white dwarf stars, which are objects of
$R_{\rm wd} \sim 10^4$\,km. My project was to try to understand how
the presence of a field of this general size might affect energy loss
from a cooling white dwarf.

I first tried to understand how a strong field might affect heat
transfer through the white dwarf's atmosphere and in the
non-degenerate outer layers. I did not make a lot of progress in these
directions, although I did realise that there could be ``cyclotron
opacity'' under certain circumstances, and that this could lead to net
circular polarisation of the emitted radiation. I also found that
sufficiently strong fields could produce strong magneto-resistivity
normal to the field in the degenerate white dwarf interior, but that
the effects of this would probably only be detectable (in the form of
obviously non-uniform surface temperature) for fields of, say, $10^9$\,G
or more. 

Then one of my Physics Department advisers, Gerald Feinberg, suggested
that I try to estimate the enhanced rate of direct cooling of the
interior of a highly magnetised white dwarf due to what we might call
synchrotron radiation of neutrinos. This effect occurs when very
occasionally the weak interaction leads to production (and immediate
loss) of a neutrino-anti-neutrino pair in place of emission of a
photon during synchrotron emission by an electron spiralling in a strong
field. The question was to understand whether this could be a
significant extra heat loss mechanism that could cause strongly
magnetic white dwarfs to cool abnormally rapidly. It turns out that
the effect does occur, but is unable to significantly affect the
cooling of even very strongly magnetised white dwarfs. I submitted my
thesis in August 1965, and successfully defended it the following
January. I published the single paper from my thesis in the Physical
Review \citep{Land67}, incidentally ensuring that it would
{\em not} be seen by any significant fraction of the stellar physics
community.

In the fall of 1965 I became an instructor at Mount Holyoke
College, where I taught physics until the summer of 1967. As I had no
significant community of collaborators or stimulators, I had some
trouble finding my next astrophysics problem, but I did become
interested in computing models of the evolution of stellar internal
magnetic fields, and eventually also in modelling the observed surface
fields of magnetic Ap stars. In the summer of 1967 I returned to New
York as a post-doctoral fellow under my old boss Woltjer, just in time to be in
close touch with the discovery of and some early work on pulsars. 

\section{Discovery}

A few months after I returned to New York City, the first pulsars were
discovered by Jocelyn Bell and announced by \citet{Hewietal68} Almost
simultaneously, Woltjer's friend Franco Pacini proposed that magnetised
neutron stars could emit radio waves \citep{Paci67}. Further pulsars began
to be discovered, and discussions were frequent among theorists
interested in this phenomenon, including Woltjer. The consensus
quickly formed that pulsars are rapidly rotating, very strongly
magnetised neutron stars \citep{Paci68a,Paci68b,Gold68}, a basic
picture which has turned out to be essentially correct. As you can
imagine, this was very exciting stuff for people interested in
supernovae, neutron stars, and magnetic fields.

Sometime around this time, a physics professor at the University of
Oregon, James Kemp, became interested in the polarisation of thermal
radiation emitted by heated objects. He predicted, and observed in
laboratory experiments that he carried out, that thermal radiation
from heated metals (i.e. materials with free electrons) in a strong
ambient magnetic field show broad band (continuum) circular
polarisation of the order of $10^{-2}$\,\% polarisation for a field of
the order of $10^5$\,G observed along the axis of the field
\citep{Kemp70,Kempetal70a}. The significance for our narrative of 
this apparently totally unrelated work will become clear later.

At the same time that this was going on, I had become very
interested in the magnetic fields of the magnetic Ap main sequence
stars. I was trying to build numerical models of the internal fields
that could be evolved, and also trying to find a surface magnetic
field distribution that would explain the variations of the mean
line-of-sight magnetic field strength $\langle B_z \rangle$ that had
been obtained by Babcock through the rotation cycle of a few Ap stars
\citep{Babc58a,Babc58b}. This second quest put me in touch with
George Preston, who had recently moved from Lick Observatory to the
Hale Observatories in Los Angeles, and had begun a
program of (photographic) polarised spectroscopy of magnetic Ap
stars. Preston invited me to visit him in February 1969, and we
found that we had a lot of interests in common. Preston loaned me some
actual polarised spectra of real magnetic Ap stars, and effectively
became my second mentor.

Then Woltjer made his second key contribution to the winding trail
that led to the discovery of white dwarf magnetism. He had a
discussion with Roger Angel, an Alfred P. Sloan Foundation Fellow
in the Columbia Physics Department. Angel was a post-doctoral fellow working in the
Columbia Radiation Lab on rocket experiments, and he wanted an
unrelated project (one that did {\em not} depend on the success of a rocket
launch) that he could work on in parallel with his rocket work. Woltjer
suggested that Angel should try to discover a magnetic field in a
white dwarf.

It was already clear, from the absence of obvious Zeeman splitting in
the spectra of the more than one hundred white dwarfs with strong line
spectra that had been observed spectroscopically at low resolution,
mainly by Jesse Greenstein \citep{EggeGree65,EggeGree67}, that fields
larger than, say, 1\,MG must be comparatively rare, so to find a
magnetic white dwarf one would need to look for fields ten or more
times weaker. This would require the development of a polarimeter to
measure circular polarisation in the spectral lines of these faint
(mostly fainter than $V \sim 12.5$) objects. 

Angel accepted the challenge immediately, and during the spring and early
summer of 1969 he designed and had the Physics Department shop build a
portable photoelectric filter polarimeter that, with interference filters,
could be used to search for and measure the circular polarisation
expected in the wings of the strong, broad Balmer lines, and  that should be
detectable even in fields of the order of $10^4 - 10^5$\,G. As the local
``expert'' on white dwarfs (but completely ignorant of everything
having to do with astronomical observing), I found myself swept into
the project.

By midsummer the polarimeter was finished and ready for testing and
use. It was a Cassegrain instrument, with a rapidly switched Pockels
cell quarter waveplate followed by a Wollaston prism, tunable interference
filters with 30\,\AA\ bandpass to isolate spectral line wings (the
central wavelengths could be shifted to the blue by tilting the
filters), and two photomultipliers, one for each output beam from the
Wollaston prism. The basic quarter waveplate--Wollaston combination sorted
the incoming light into two beams, each with intensity proportional to
one of the two states of circular polarisation. The fast and slow
axes of the Pockels cell were controlled by an applied high voltage,
switched rapidly to allow each photomultiplier channel to measure the
intensity of both polarisation states independently, using a pair of
gated scalers for each channel. The instrument is described briefly by
\citet{AngeLand70a}.

Angel arranged for two observing runs with the istrument at McDonald
Observatory in Texas (by purchase, I think -- time on the 82-inch was
available at \$450 per night, and at \$750 on the 107-inch). We had
about a week on the 36-inch in August, and a few nights on the 82-inch
in September. During these runs Angel and I surveyed a total of nine
bright DA white dwarfs. Such white dwarfs show only spectral lines of
hydrogen, and because of the very high surface gravity of these stars
($g \sim 10^8$\,cm\,s$^{-2}$) their Balmer lines have full widths of
the order of 100--200\,\AA. We detected no polarisation in any of the white
dwarfs we observed, but were able to put upper limits to possible
line-of-sight fields of between 10$^4$ and 10$^5$\,G
\citep{AngeLand70a}. We almost immediately applied for more telescope
time, this time at Kitt Peak National Observatory near Tucson, AZ. We
were awarded 3 nights on the 84-inch and a week on a 36-inch telescope
between June 26 and July 5, 1970. 

In the meantime, two key events on our winding trail occurred. 

Sometime in autumn of 1969, George Preston made a visit to the
Astronomy Department at Columbia University. I have a vague
recollection that this visit lasted for a couple of weeks, but I am not
sure. In any case, he heard all about the search that Angel and I were
carrying out for magnetic white dwarfs. One consequence of the visit
was that Preston realised that the fields in question (of the order of
1\,MG) are large enough that, in hydrogen, the quadratic Zeeman effect
leads to significant wavelength shifts that increase with increasing
principal quantum number $n$ of the upper level, and so are larger for
the higher members of the Balmer series. Preston was aware of a
careful study by \citet{GreeTrim67} of radial velocities of white dwarfs
using several Balmer lines in each star. He discussed the
uncertainties of these measurements at length with Greenstein, and
concluded that a field larger than about 5\,10$^5$\,G would have
alerted Greenstein \& Trimble to a problem with the
measurements. Preston's conclusion was that none of the 60-some WDs
measured by Greenstein \& Trimble has a magnetic field of more than
about 5\,10$^5$\,G, and thus that searches for white dwarf fields by
polarimetry needed to be sensitive to fields of order 10$^5$\,G or less. 

But the most important consequence, for our narrative, of Preston's
interest in the magnetic fields of white dwarf stars occurred later
that year, when he visited the University of Oregon. He met James
Kemp, the Oregon professor interested in the polarisation of thermal
radiation whom I mentioned above. Kemp understood from his discussions
with Preston that people were thinking of the possible occurrence of
fields in (some) white dwarfs of the order of 1\,MG, and that if such
fields existed, detecting broadband circular polarisation could be a
method of identifying these fields. Detecting the first MG field in a
white dwarf would beautifully confirm the basic physical effect the
Kemp had been studying, and would also be a very exciting discovery in
its own right. Kemp immediately started work to adapt his laboratory
polarimeter to measure the polarisation of starlight on the new
24-inch telescope of the university's Pine Mountain Observatory, and
soon began a programme of observation of the brightest (mostly DA)
white dwarfs.

In early 1970, Kemp visited Columbia University, and he and I
discussed his work and his search for broadband circular polarisation
in white dwarfs. I explained to him that the programme that Angel and
I were carrying out, using spectral lines, was a more sensitive method
of detecting weak fields than his broad-band searches. However, I
suggested that his method would be a very powerful way to search for
large fields in DC white dwarfs. These are white dwarfs showing {\em
no} spectral lines, and hence white dwarfs for which no {\em a priori}
upper limits to magnetic fields existed from the absence of the Zeeman
effect in normal flux spectra. I gave Kemp finding charts for several
bright DC stars, and in particular I urged him to observe the very
strange white dwarf Grw+70$^\circ$\,8247 in which the unique weak
spectral absorption features (the ``Minkowski bands'') were still
completely unexplained, as this white dwarf seemed to me to be the
most likely star to show Kemp's predicted effect.

In June of 1970, Angel and I took Angel's polarimeter to Kitt Peak
Observatory to continue our searches for weak fields in DA stars. On
June 30, Kemp called Angel to tell us that Kemp's Pine Mountain
polarimetric observations had twice detected a strong signal of
circular polarisation in Grw+70$^\circ$\,8247. He asked us to confirm
his result. We removed the narrow band filters from our polarimeter,
replaced tham with broadband glass filters, and observed
Grw+70$^\circ$\,8247 that night. Our observations fully confirmed Kemp's
discovery of broadband circular polarisation. 

Grw+70$^\circ$\,8247 thus became the first magnetic white dwarf ever
discovered. Kemp initially estimated the field at roughly 10$^7$\,G
(we now know that the field is more nearly 3\,10$^8$\,G). The event
was sufficiently exciting that the Astrophysical Journal Letters put a
news embargo on the paper \citep{Kempetal70b} until it was published,
as Science and Nature magazines often do for today's hot new results.
The discovery of the huge field of Grw+70$^\circ$\,8247 became a ``new
discovery'' news item in both Science and Nature magazines, and even
made it into the pages of Time Magazine.

\section{Sequel}

Angel and I followed up this exciting discovery vigourously. Armed
with great first results, we were able to get more telescope
time, and rather quickly found three more bright,
circularly polarised magnetic white dwarfs: G195-19
\citep{AngeLand71a,AngeLand71b}, G99-37 \citep{LandAnge71}, and
G99-47 \citep{AngeLand72}. Kemp and his group also discovered another
circularly polarised white dwarf, GD229 \citep{Swedetal74}. All of
these stars were discovered because of detectable continuum circular
polarisation. It was only after several years that it was diswcovered
that one of these stars, G99-47, actually shows normal Zeeman
splitting of H$\alpha$ \citep{Liebetal75}. Zeeman splitting in flux
spectra has, of course, been subsequently observed in numerous fainter
white dwarfs

Gradually, at a discovery rate of one or two new magnetic white dwarfs
per year (up until about 2003, when the SDSS spectroscopic follow-up
survey began to uncover tens and even hundreds of new magnetic white
dwarfs with fields of 2--80\,MG through Zeeman splitting), new fields
were found, until now it is clear that 10\% or more of white dwarfs have
fields, with the distribution of fields covering the huge range of
strengths from a couple of kG up to 1000\,MG
\citep[e.g.][]{LandBagn19}. Magnetism is now a well-established
branch of white dwarf physics: see Stefano Bagnulo's talk at this
meeting.

The main point of this talk is that the discovery of the first
magnetic white dwarf happened as a result of an (unlikely?) series of
ideas and conversations. The initial possibility of finding MG fields
was proposed by Blackett, and many years later by Lo Woltjer. Woltjer
suggested to Roger Angel that a search for large fields should be
initiated. Angel took up the challenge, quite incidentally including
me in the project. Our search was initially unsuccessful, but (because
of his interaction with Woltjer and me) George Preston heard about our
search, and (improbably) passed the idea of MG fields in white dwarfs
to physicist Jim Kemp at the University of Oregon, who decided to
search for {\em continuum} polarisation in white dwarfs. He was also
unsuccessful until he talked with me, and I explained to him that he
should be looking at the DC stars, the one class of white dwarfs for
which no strong prior upper limits on possible magnetic fields are known
from optical spectroscopy. In particular, I urged Kemp to observe
Grw+70$^\circ$\,8247, which spectacularly displays the circular
polarisation predicted by him.  And so, finally, in 1970, the first
magnetic white dwarf was found.

\section{Postscript: why didn't Jesse Greenstein discover the first magnetic white dwarf?}

There is one further aspect of this story that is intereting to
consider. We now know that magnetic fields occur in at least about
10\% of white dwarfs
\citep[e.g.][]{Jordetal07,KawkVenn14,LandBagn19}. Spectrocsopic observation of
white dwarfs, particularly by Jesse Greenstein using the 200-inch
telescope, was racing ahead \citep{EggeGree65,EggeGree67,Gree69}; by
1970 more than 250 white dwarfs had been observed and classified
spectroscopically at low resolving power ($R \sim 10^3$). It is
interesting to ask why the first magnetic white dwarf was not
discovered by observation of magnetic line splitting in the available
optical spectra during the 1960s.

The answer is probably buried in the actual details of the instruments
and their limitations. The 10\% or more of white dwarfs
with magnetic fields are now known to have fields ranging from a few
kG to almost 1000\,MG, a range in strength of 5 orders of magnitude,
with the fields roughly uniformly distributed over this full range at
a rate of perhaps 2 or 3\% occurence per dex of field strength. Now in
fact the magnetic line splitting due to a magnetic field is only
really obvious in low resolution spectra for field strengths above
about 2\,MG, and fields above about 20\,MG lead to spectra that, due
to the different shifts of the many sublevels of each level of H, are
sufficiently complex to be quite difficult to interpret, if indeed the
many weak lines were detected at all. So the first obstacle was that
in practice only about a fifth of the possible fields might have been
easily recognised by a spectroscopist. This reduces the probability
of finding a visibly magnetic white dwarf to roughly 2 or 3\% of a
sample.

But there were still further difficulties. Of the presently known
magnetic white dwarfs, only about 25 or 30 are brighter than $V = 15$
(and thus could be observed during the 1960s with relatively high
$S/N)$. Only two of these brightest magnetic white dwarfs, G99-47 and
Feige 7, have fields in the range easily detected by visual inspection
of optical spectra. Both were in fact observed by Greenstein, who
classified both as DC white dwarfs, in one case because the only
visible spectral line lies outside the blue spectral window used for
almost all spectroscopy during the 1960s, and in the other case
because the numerous weak blue spectral lines simply were not obvious
in the low $S/N$ spectrum.

In contrast, fields large enough to generate an easily detected
circular polarisation signal of, say, $V/I \approx 0.5$\% are present
in perhaps 5\% of white dwarfs. Furthermore, the use of photoelectric
detectors, with their far higher detective quantum efficiency compared
to the photographic and image tube technology used for spectroscopy
during the 1960s, together with the fact that the full optical
spectrum contributes to massively increase the $S/N$ of the single
polarimetric broad-band observation, meant that searches for
circularly polarised white dwarfs could be carried out relatively
rapidly even on telescopes considerably smaller than the 200-inch. The
first white dwarf field, after all, was discovered using a 24-inch
telescope and confirmed on a 36-inch telescope
\citep{Kempetal70b}. Kemp's new broadband circular polarimetric search method,
combined with the enormous increase in overall spectral efficiency due
to multiplexing and greatly increased detective quantum efficiency
compared to spectroscopy, turned out to be the best available search
method before the arrival of CCD spectroscopy.

\acknowledgements{I have greatly appreciated working with a large
group of outstanding collaborators, who have hugely enriched my
career and my personal life. I am enormously grateful to Aaron Sigut,
David Bohlender, Gregg Wade, and Sarah Landstreet for organising this
wonderful conference, to Henry Leparskas and Michel Debruyne for terrific
support and innumerable great photos, and to Zden\v{e}k
Mikul\'a\v{s}ek for many happy hours playing our clarinets together. }

\bibliographystyle{ptapap}
\bibliography{landstreet}

\end{document}